\documentclass[11pt]{article}
\usepackage{epic, eepic}
\usepackage{units}
\usepackage{amsthm, amsmath, amssymb,amscd}
\usepackage[all]{xy}
\usepackage{mathrsfs}
\usepackage{graphicx}
\usepackage{epstopdf}
\usepackage{enumerate}

\newtheorem{thm}{Theorem}[subsection]

\newtheorem{proposition}[thm]{Proposition}

\newtheorem*{remark}{Remark}

\usepackage{fullpage}

\begin{document}
\title{Parametric analysis of RNA folding} 
\author{Valerie Hower, Christine E. Heitsch}
\date{}
\maketitle
\begin{abstract}
We extend recent methods for parametric sequence alignment to the parameter space for scoring RNA folds.  This involves the construction of an RNA polytope.  A vertex of this polytope corresponds to RNA secondary structures with common branching.  We use this polytope and its normal fan to study the effect of varying three parameters in the free energy model that are not determined experimentally.  Our results indicate that variation of these specific parameters does not have a dramatic effect on the structures predicted by the free energy model.  We additionally map a collection of known RNA secondary structures to the RNA polytope.  

\end{abstract}

\section{Introduction}
Determining the structure of RNA molecules remains a fundamental scientific challenge, since current methods cannot always identify the ``correct'' fold from the large number of possible configurations.  A common method for predicting the secondary structure of a single RNA molecule, termed the \emph{thermodynamic model}, involves free energy minimization \cite{mathews-turner-06, zuker-00, zuker-mathews-turner-99}.  Extensions to this approach, such as suboptimal structure
prediction and partition function calculations, still depend on the parameters from the thermodynamic model to score possible secondary structures.  The free energy of a secondary structure is calculated by scoring substructures according to a set of parameters---most of which are determined experimentally (see \cite{SantaLucia} for a review).  A dynamic programming algorithm, used in software packages like mfold \cite{mathews-etal-99, zuker-03}, computes the minimal free energy as well as the optimal secondary structure(s) \cite{mathews-etal-04}.

In this work, we address variation in the parameter space for scoring secondary structures, focusing on three parameters from the multi-branch loop energy function that are not based on measurement.  Specifically, we address the following questions. What is the geometry of the parameter space for scoring RNA folds and how does this geometry relate back to the biology?  How sensitive is the thermodynamic model to variation of the ad-hoc multi-branch loop parameters?  We answer these questions using geometric combinatorics.  We find that variation of the multi-branch loop parameters has a smaller effect than the change in the parameter space coming from improved measurement.  Moreover, regardless of the choice of multi-branch loop parameters used in the current version of the thermodynamic model, the minimal energy structures have a low degree of branching.        

Our results are achieved by applying techniques from geometric combinatorics to give a parametric analysis of RNA folding.  We construct an RNA polytope whose vertices correspond to sets of secondary structures with common branching. Its normal fan subdivides the parameter space so that the parameters lying in the same cone give the same minimal free energy structures.  These approaches have been used recently in parametric sequence alignment  \cite{align1, align3, align2} and for more general hidden Markov models \cite{recombination, hmm}.  There is also earlier polyhedral work on parametric sequence alignment \cite{align00, align0} and related work on secondary structure comparison \cite{wang-zhao-03} and sequence/structure alignment \cite{lenhof-reinert-vingron-98}.  We additionally make comparisons with biological structures, and this work supports our theoretical results.
\section{Background}
\subsection{Plane trees and RNA folding}
\begin{figure}[htdp]
\begin{center}
{\bf A)}\includegraphics[width=2.75in]{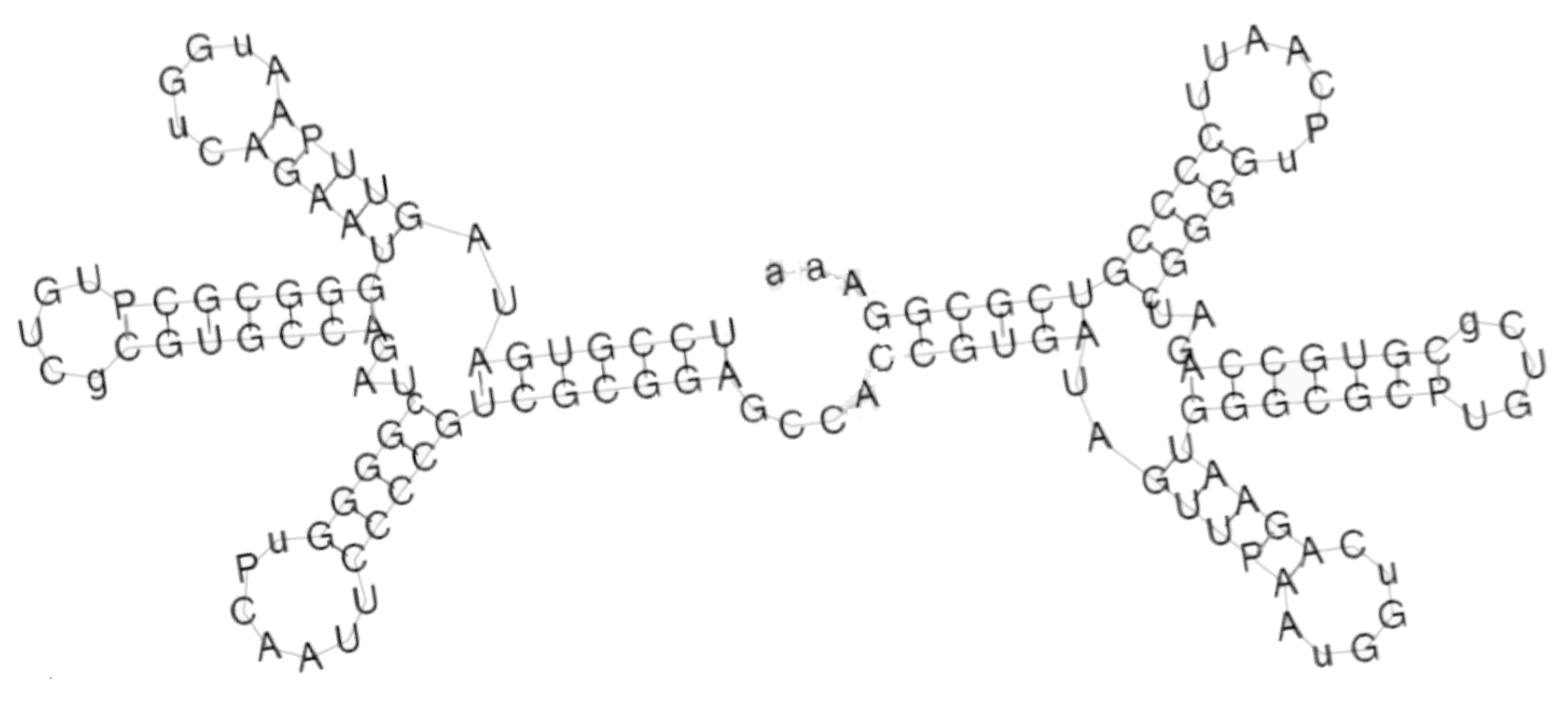} {\bf B)} \includegraphics[width=1.7in]{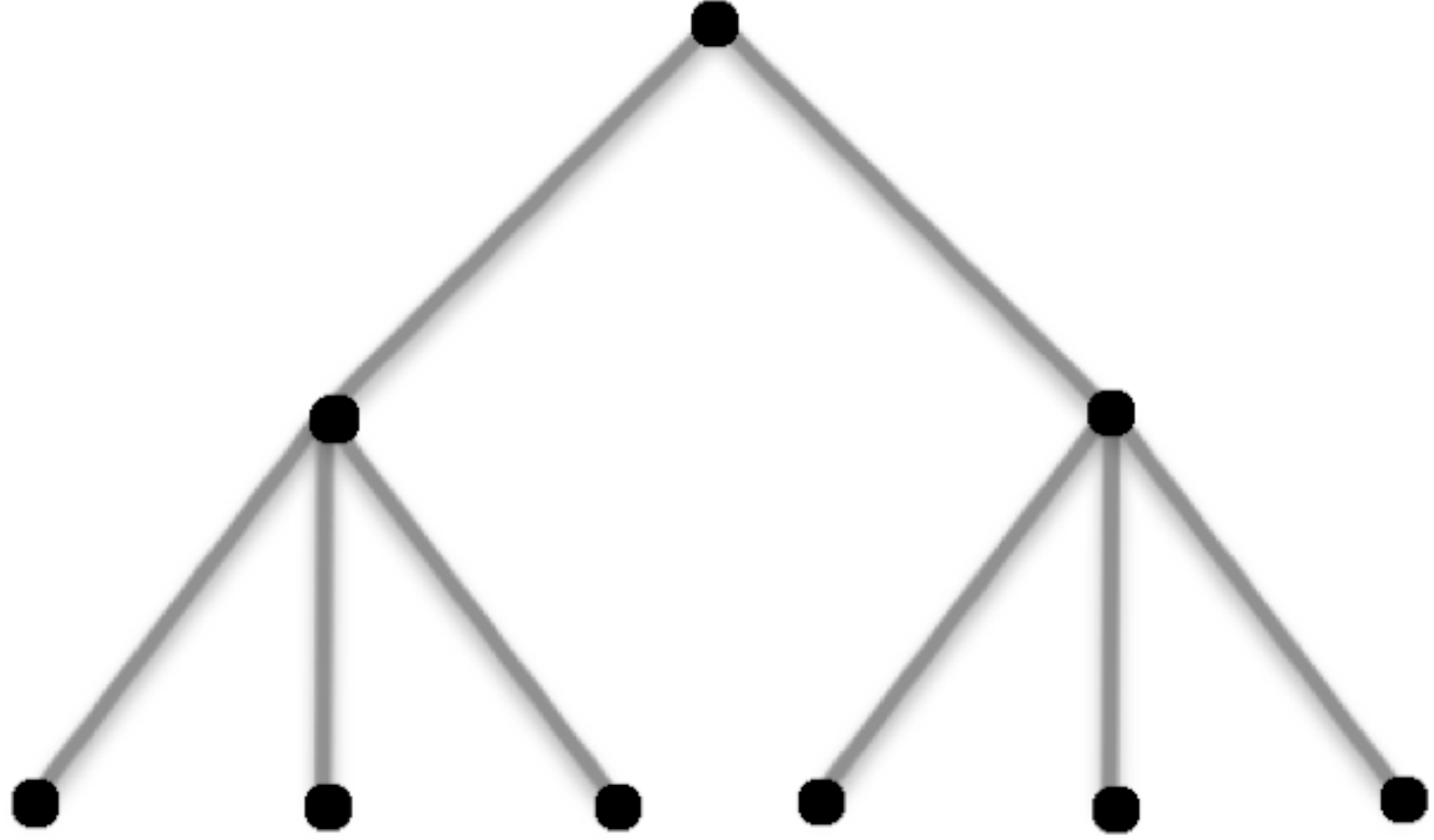}
\caption{Secondary structures as rooted plane trees}
\label{figure:ST}
\end{center}
\end{figure}

We use a simplified model of RNA folding in which a secondary structure $S$ is represented by a rooted plane tree $T=T(S)$.  Single-stranded RNA sequences fold into molecular structures.  One step in this folding process is the formation of Watson-Crick and also G-U base pairs.  The set of (nested) base pairs determines the secondary structure of an RNA sequence.  As illustrated in Figure \ref{figure:ST}, a secondary structure has two basic types of substructures---runs of stacked base pairs which are called \emph{helices} and the single-stranded regions known as \emph{loops}.   Every component of a
secondary structure is given an associated free energy score by the thermodynamic model.  To a first approximation, the score of a loop is determined its \emph{degree}---the number of base pairs contained in the loop.  There are different energy functions for the external loop, hairpin loops, which have degree $1$, bulge/internal loops with degree $2$, and multi-branch loops with degree greater than $2$.  Suppose $L$ is multi-branch loop, then the free energy of $L$ is 
\begin{equation} \label{30}
E(L)= a+bn_1+cn_2+q,
\end{equation}
where $n_1$ is the number of single-stranded bases in $L$, $n_2$ is the number of helices in $L$, $q$ is the sum of the single-base stacking energies in $L$, and $a,b,c$ are the parameters for offset, free base, and helix penalties, respectively \cite{zuker-mathews-turner-99}.  In this work, our analysis is primarily focused on the three parameters $a,b,$ and $c$ from this function since they are not experimentally determined.   Our results are obtained by considering rooted plane trees as a simplified model of RNA folding.

Plane trees have been used to enumerate possible RNA secondary structures \cite{schmitt-waterman-94} and also to compare them \cite{le-nussinov-maizel-89,
shapiro-zhang-90} for some time now.  The interaction between combinatorics and RNA folding has continued to develop over the last 20 years, including using trees as more abstract representations of RNA folding, for instance in \cite{gan-pasquali-schlick-03} and related work as well as
in \cite{bakhtin-heitsch-08, bakhtin-heitsch-09,heitsch-unpub}.  A \emph{rooted plane tree} (also called \emph{plane tree} or \emph{ordered tree} \cite{plane1980, stanley-99})  is a tree with a specified root vertex and such that the subtrees of any given vertex are ordered.  This ordering comes from the $5^{\prime}\rightarrow 3^{\prime}$ linear arrangement of the RNA sequence.  Plane trees with $n$ edges are one of the many combinatorial objects counted by the Catalan numbers  \begin{equation} \label{catalan}
C_n=\frac{1}{n+1}{2n\choose n} .
\end{equation}
To obtain $T$, we assign the root vertex to the exterior loop of $S$ and the non-root vertices of $T$ correspond to the remaining loops in $S$.  Two vertices in $T$ share an edge when their  loops in $S$ are connected by helices.  As an example, we give a secondary structure in Figure \ref{figure:ST}{A} together with its associated plane tree in Figure \ref{figure:ST}{B}.  Technically, a secondary structure $S$ must be free of pseudoknots in order to construct $T$.  While pseudoknots do occur in secondary structures, the thermodynamic model cannot predict them and moreover one can create a nested, pseudoknot-free structure from a given fold in several ways---some of which are in \cite{remove} and our approach is described the Materials and Methods section.
Given a plane tree $T$ with $n$ edges, we write $r$ for the degree of the root vertex and for $0\leq k \leq n$, $d_k$ is the number of non-root vertices with $k$ children.  Thus, $d_k$ gives the number of non-root vertices in $T$ with degree $k+1$, and this is the number of loops in $S$ with $k+1$ branches.  To assign an energy to a plane tree, we assign weights to the vertices, based on the down degree of the vertex.  In terms of secondary structures, we are assigning the same energy to each type of loop in the fold.  This is a simplification of the scoring for the thermodynamic model, in which the energy of a structure is the sum of the energies of the loops.  If $T$ is a plane free with $n$ edges, the free energy of $T$ is written as 
\begin{eqnarray*}
E(T)&& =  a_3r+a_0d_0+a_1d_1+ \sum_{k=2}^n[c_2+a_2(k+1)]d_k \\
&&= a_3r+a_0d_0+a_1d_1+ (c_2+2a_2)\sum_{k=2}^nd_k +  a_2\sum_{k=2}^n(k-1)d_k \\
&&= a_3r+a_0d_0+a_1d_1+(c_2+2a_2)(n-d_0-d_1)+a_2(d_0-r)  \\
&&=(c_2+2a_2)n +(a_3-a_2)r + (a_0-c_2-a_2)d_0 +(a_1-c_2-2a_2)d_1,
\end{eqnarray*}
where we have used the relations $$\sum_{k=2}^nd_k=n-d_0-d_1\quad \mbox{and} \quad \sum_{k=2}^n(k-1)d_k=d_0-r$$ that hold for all plane trees \cite{stanley-99}.  To minimize free energy, we must minimize $E(T)$ over the space of all plane trees.  Since this space is infinite, we will typically think of $n$ as being fixed but arbitrary and minimize the free energy function over the finite space of plane trees with $n$ edges.    For a given set of parameters $a_0, a_1, a_2, a_3, c_2$, this is equivalent to minimizing the following inner product 
\begin{equation}\label{eprime}E^{\prime}(T)=(\theta_2, \theta_3,\theta_4)\cdot (r,d_0,d_1)\end{equation} where 
$$\begin{array}{lll}
  \theta_2= a_3-a_2 & \theta_3=a_0-c_2-a_2 &\theta_4=a_1-c_2-2a_2. \\
\end{array} $$

\subsection{Geometric combinatorics}  In this section, we present some basic definitions in geometric combinatorics.  We refer the reader to \cite{grunbaum, ziegler} for a more detailed treatment.
 A set $U\subset \mathbb{R}^d$ is \emph{convex} if for any two points $x,y\in U$, the line segment connecting $x$ and $y$ is contained in $U$, that is $\{\alpha x +(1-\alpha )y \, \, | \, \, 0\leq \alpha \leq 1\} \subset U$.  For any subset $U$ of $\mathbb{R}^d$, the \emph{convex hull} of $U$, written $\mathrm{conv}U$, is the intersection of all convex sets that contain $U$.
A \emph{lattice polytope} $\Delta \subset \mathbb{R}^d$ is the convex hull of a finite collection of lattice points: $\Delta=\mathrm{conv}\mathcal{A}$, where $\mathcal{A}=\{y_1, y_2, y_3, \cdots , y_r \}\subset \mathbb{Z}^d$.

Any lattice polytope $\Delta$ is characterized by a finite collection of defining inequalities 
\begin{equation}\label{defining}
\{c_i\cdot x \ge b_i\}_{i\in I}\quad\mbox{where} \quad c_i\in\mathbb{Z}^d, \, x\in\Delta, \mbox{ and } b_i\in \mathbb{Z}.\end{equation}  
A \emph{face} $F$ of $\Delta$ is a subset defined by setting some of the defining inequalities to equality, i.e.
$$F=\left\{ x\in \Delta \, \left| \,\begin{array}{c}c_{i_1}\cdot x = b_{i_1}\\ c_{i_2}\cdot x = b_{i_2} \\ \cdots \\ c_{i_k}\cdot x = b_{i_k}\end{array}\right. \right\},$$ and the dimension of $F$ is the dimension of its affine span.  The \emph{vertices} of $\Delta$ are the $0$-dimensional faces while the \emph{facets} have dimension $\mathrm{dim}\Delta - 1$.  The \emph{boundary} of $\Delta$, written $\partial \Delta$ is the union of all faces of $\Delta$ of dimensions $0,1,2, \cdots , \mathrm{dim}\Delta -1$.   

A \emph{convex polyhedral cone} $\sigma$ is the positive hull of a finite collection of lattice points in $\mathbb{Z}^d$: $\sigma=\{t_1z_1+t_2z_2+\cdots +t_sz_s \, \, | \, \, t_i\ge 0, z_i \in \mathbb{Z}^d\}$, and we write $\sigma=\langle z_1, z_2, \cdots z_s\rangle $.  Associated to each lattice polytope $\Delta$ is its normal fan $\mathcal{N}(\Delta)$ that is a collection of cones and subdivides $\mathbb{R}^d$.  
The \emph{rays} ($1$-dimensional cones)  of $\mathcal{N}(\Delta)$ are of the form $\langle c_i \rangle$ for $i\in I$ in \eqref{defining}.  Moreover, the cones $\sigma \in \mathcal{N}(\Delta)$ are in one-to-one correspondence with faces $F$ of $\Delta$: 
\begin{equation}\label{coneeq}
 \sigma_F=\{v\in \mathbb{R}^d \, \, | \, \, u\cdot v \leq x\cdot v \quad \forall u\in F,\, \forall x\in \Delta\}.
\end{equation}

\begin{figure}[htbp]
\begin{center}
{\bf A)}\includegraphics[width=1.5in]{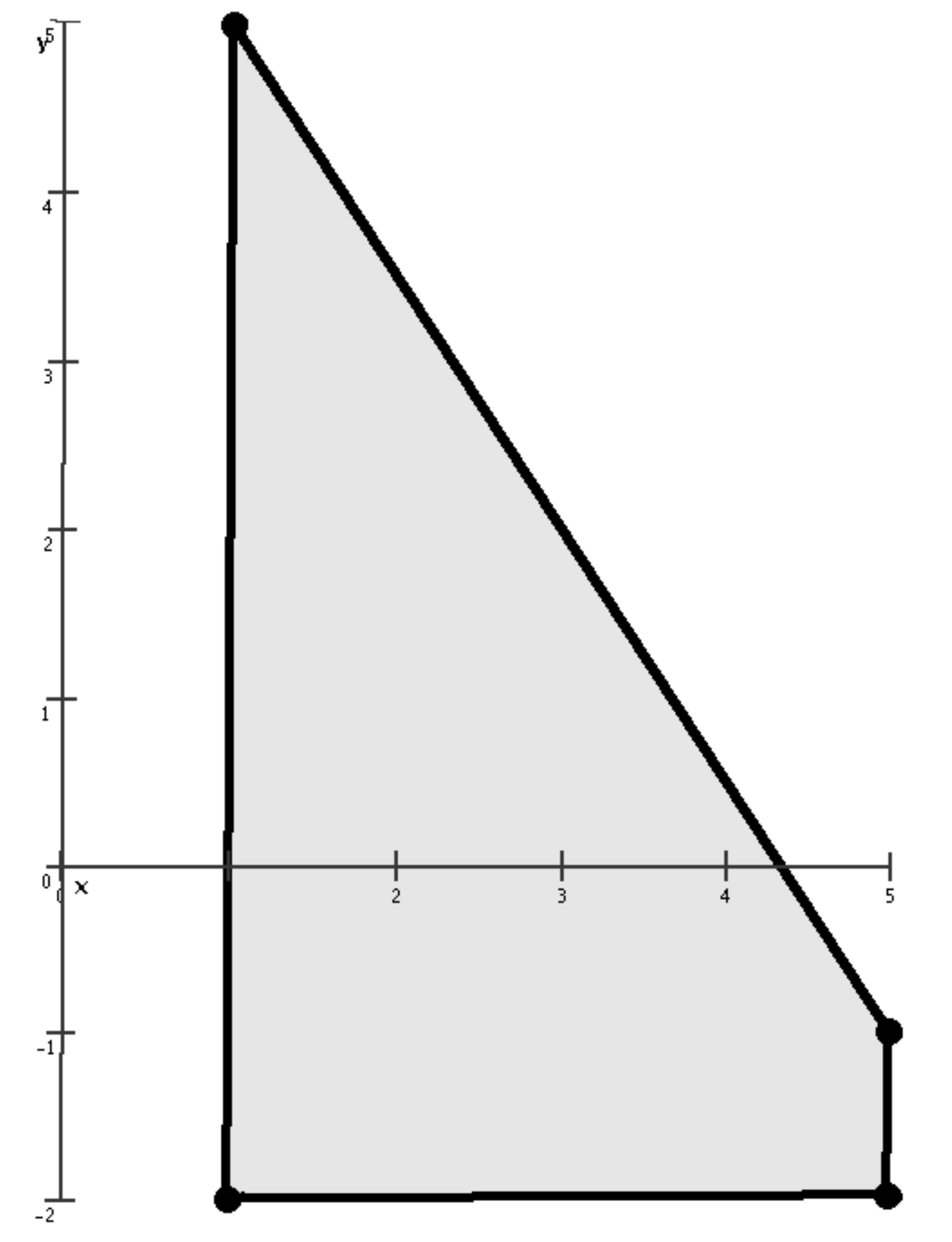} {\bf B)} \includegraphics[width=2.2in]{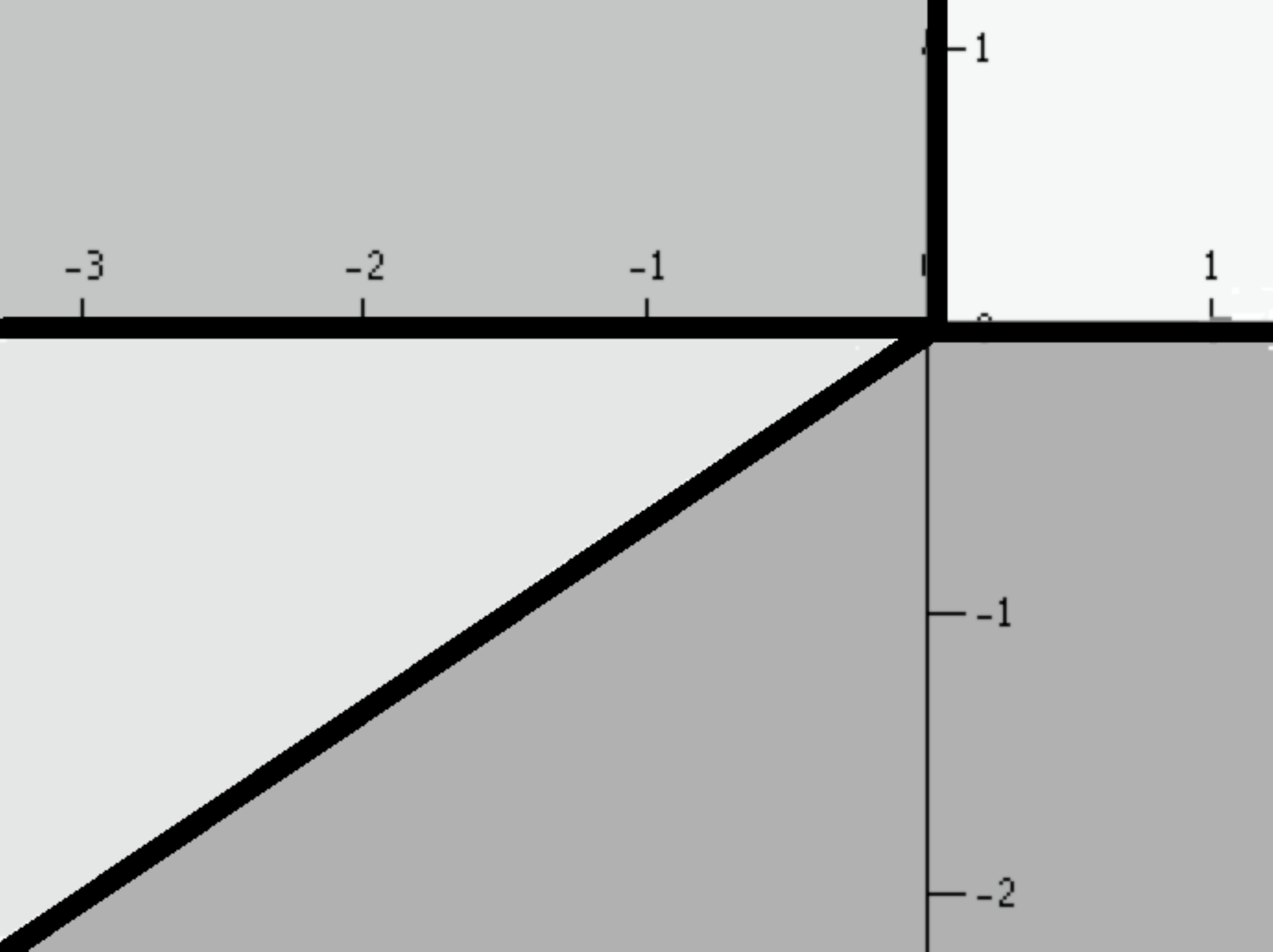}
\caption{A $2$-dimensional polytope $\Delta$ (A) and its normal fan $\mathcal{N}(\Delta)$ (B) }
\label{figure:polygon} 
\end{center}
\end{figure}

Note that $\mathrm{dim}\sigma_F=\mathrm{dim}\Delta - \mathrm{dim}F$.  In terms of minimization, equation \eqref{coneeq} states that the points in $F$ are minimizers of the dot product for vectors in $\sigma_F$, among all points in $\Delta$.  As an example of the above concepts, we give a $2$-dimensional polytope $\Delta$ in Figure \ref{figure:polygon}{A} and its normal fan $\mathcal{N}(\Delta)$ in Figure \ref{figure:polygon}{B}.  The four vertices of $\Delta$ correspond to the four $2$-dimensional cones in $\mathcal{N}(\Delta)$, and the four facets of $\Delta$ correspond to the four rays of $\mathcal{N}(\Delta)$.

\section{Results}
\subsection{Plane trees that minimize energy}
Fixing $n\ge 5$, the possible count vectors $(r,d_0,d_1)$ of plane trees are classified by the second author \cite{heitsch-unpub} and fall into one of four classes, as listed in Table \ref{table:ineq} with $r,d_0,d_1\geq 0$ in all cases. 
\begin{table}[htbp] 
\centering     
\begin{tabular}{cc c c} 
\hline\hline                  
& Set of inequalities &  Vertices for $n$ even & Vertices for $n$ odd \\ [0.5ex] \hline
(A)& $\begin{matrix} r=1\\ d_0=1\\  d_1=n-1 \end{matrix}$ &$\{(1, 1, n-1)\}$ &$\{(1, 1, n-1)\}$  \\ \\
(B)& $\begin{matrix} r=1 \\  2\leq d_0\leq n \\n-2d_0+1\leq d_1\\ d_1 \leq n-d_0-1 \end{matrix} $&$\left\{\begin{matrix} (1,2,n-3), (1,\frac{n+2}{2}, 0), \\ (1, \frac{n}{2},1),  (1,n-1,0) \end{matrix}\right\}$ &$\left\{\begin{matrix} (1,2,n-3), (1,\frac{n+1}{2}, 0), \\ (1,n-1,0) \end{matrix} \right\}$\\ \\
(C)& $\begin{matrix}  r=d_0\\ 2\leq d_0 \leq n \\ d_1=n-d_0 \end{matrix}   $  &$\{(2,2,n-2),(n,n,0) \}$&$\{(2,2,n-2),(n,n,0)\}$ \\ \\
(D)& $\begin{matrix} 2\leq r\\ r \leq 2d_0-n+d_1 \\ 3\leq d_0 \leq n-1 \\  n-2d_0+2\leq d_1\\d_1\leq n-d_0-1 
      \end{matrix}$ & $\left\{\begin{matrix} (2,n-1,0), (n-2,n-1,0), \\ (2,3,n-4), (2, \frac{n+2}{2},0)\end{matrix}\right\} $ &$\left\{\begin{matrix}(2,n-1,0), (n-2,n-1,0), \\  (2,\frac{n+3}{2},0), (3,\frac{n+3}{2},0), \\  (2,3,n-4), (2,\frac{n+1}{2},1)\end{matrix}\right\}$ \\ 
\hline      
\end{tabular} 
\caption{Sets of inequalities and corresponding vertices for plane trees}  
\label{table:ineq}  
\end{table}
 Since $r,d_0,d_1$ must all be integers, the vertices in Table \ref{table:ineq}{B} or Table \ref{table:ineq}{D} differ depending on whether or not $n$ is even or odd.  We want to minimize the linear energy function over this point set (which includes count vectors from all four cases), and hence we let $P_n$ be the convex hull of the union of the four polytopes listed in Table \ref{table:ineq}.  Regardless of our choice of energy parameters, a minimum energy plane tree with $n$ edges will occur at a vertex of $P_n$.  The following proposition describes the vertices of $P_n$.
\begin{proposition} \label{psin}
 Define $\Psi_n$ as follows.  $$\Psi_n:=\left\{\begin{array}{ll} \mathrm{conv}\{(1,\frac{n+1}{2}, 0), (1,n-1,0), ( 1, 1, n-1),(n,n,0) \}&  n\mbox{ odd} \\ \mathrm{conv}\{(1,\frac{n+2}{2}, 0), (1, \frac{n}{2},1), (2, \frac{n+2}{2},0) , (1,n-1,0), ( 1, 1, n-1),(n,n,0) \}&  n\mbox{ even} \end{array} \right.$$  Then $\Psi_n=P_n $ for $n\ge 5$.
\end{proposition}
\proof Clearly $\Psi_n \subset P_n$ and hence we'll show each lattice point of $P_n$ in Table \ref{table:ineq} is contained in $\Psi_n$.  The normal fan of $\Psi_n$ has rays
$$ \begin{array}{ll} \{(-1,2,1),(1,0,0), (1,1-n,2-n), (0,0,1) \} & n \mbox{ odd} \\
 \{(-1,2,1),(1,0,0), (1,1-n,2-n), (0,0,1), (0,1,1) \} & n \mbox{ even} 
\end{array}
$$ 
Moreover, for each lattice point $t=(r,d_0,d_1)$ in Table \ref{table:ineq}, one can verify that $t$ satisfies the defining inequalities of $\Psi_n$:
\begin{eqnarray*}
 &(r,d_0,d_1)\cdot (-1,2,1)& \ge n  \\
&(r,d_0,d_1)\cdot (1,0,0)& \ge 1 \\
& (r,d_0,d_1)\cdot (1,1-n,2-n)& \ge 2n-n^2 \\
&  (r,d_0,d_1)\cdot (0,0,1)& \ge 0 
\end{eqnarray*}
and for $n$ even we additionally have
\begin{eqnarray*}
 &(r,d_0,d_1)\cdot (0,1,1)& \ge \frac{n+2}{2}.  
 \end{eqnarray*}
This gives $P_n\subset \Psi_n$, and we have equality. \qed 

In the sequel, we will primarily focus on the rational tetrahedron $$\Delta_n:=\mathrm{conv}\{\textstyle{(1,\frac{n+1}{2}, 0)}, (1,n-1,0), ( 1, 1, n-1),(n,n,0) \}$$ regardless of whether $n$ is even or odd.  There are many reasons for this.   First, asymptotically, there is no difference between $P_n$ and $\Delta_n$ for $n$ even.   The normal fan $\mathcal{N}(P_n)$ is obtained from $\mathcal{N}(\Delta_n)$ by adding a single ray and subdividing the full dimensional cone $\sigma=\langle (1, 0, 0), (0, 0, 1), (-1, 2, 1)  \rangle$ corresponding to the vertex $(1,\frac{n+1}{2},0)$.  Thus, when $n$ is even, the parameters giving $(1,1,n-1), (1,n-1,0),$ or $(n,n,0)$ the minimal energy are the same regardless of whether we use the subdivision of $\mathbb{R}^3$ determined by $\mathcal{N}(P_n)$ or that determined by $\mathcal{N}(\Delta_n)$.  Moreover, the parameters in $\sigma$ will yield $ (1,\frac{n+2}{2}, 0), (1, \frac{n}{2},1),$ or $(2, \frac{n+2}{2},0)$ as minimal, and the trees corresponding to these three count vectors are all similar, as discussed in Proposition \ref{vertbranch}.   
\subsection{Lattice points in $\partial P_n$}\label{boundsec}
Suppose $S$ is a secondary structure whose plane tree has count vector $(r, d_0, d_1)$.  If $(r,d_0,d_1)\in \mathrm{int}P_n$ then there is no choice of parameters that can make $S$ have minimal free energy.  Conversely, if $(r,d_0,d_1)\in \mathrm{int}F$ for some face $F$ of $P_n$, then any parameter vector in the cone $\sigma_F \subset \mathcal{N}(P_n)$ yields $S$ with minimal energy.  We thus want to determine the count vectors lying on $\partial P_n$.

All four sets of inequalities in Table \ref{table:ineq} intersect $\partial P_n$.  Let $Q_A$, $Q_B$, $Q_C$, and $Q_D$ be the polyhedra described in Table \ref{table:ineq}{A}, \ref{table:ineq}{B}, \ref{table:ineq}{C}, and \ref{table:ineq}{D}, respectively.  Then, $Q_A,Q_B,Q_C\subset \partial P_n$ and
\begin{eqnarray*}
&Q_A&=\{(1,1,n-1)\} \\
&(Q_A\cup Q_B)\cap\mathbb{Z}^3& = \mathrm{conv}\{(1,n-1,0),(1,1,n-1),(1,\textstyle{\frac{n+1}{2}},0)\} \cap \mathbb{Z}^3\\
&(Q_A\cup Q_C)\cap \mathbb{Z}^3& = \mathrm{conv}\{(1,1,n-1),(n,n,0)\} \cap \mathbb{Z}^3. \end{eqnarray*}
Since $Q_D$ is $3$-dimensional, it cannot be contained in the boundary of $P_n$.  We do, however, have
\begin{equation}\label{qd}(Q_D\cap \partial P_n)\cap \mathbb{Z}^3= (\mathrm{int}E_1\cup \mathrm{int}F_1 \cup \mathrm{int}F_2)\cap \mathbb{Z}^3, \end{equation}
where $E_1=\mathrm{conv}\{(n,n,0), (1,\frac{n+1}{2},0)\}$, $F_1=\mathrm{conv}\{(n,n,0),(1,\frac{n+1}{2},0), (1,1,n-1)\}$, and  $F_2=\mathrm{conv}\{(n,n,0),(1,\frac{n+1}{2},0), (1,n-1,0)\}$.  Equation \eqref{qd} follows from counting lattice points in the objects on the left and right hand sides of the equation using the same technique as in Proposition \ref{pick}.  The plane trees defined in Table \ref{table:ineq}{D} that lie on $\partial P_n$ satisfy $d_1=0$ or $r= 2d_0-n+d_1$.  Their associated secondary structures either have no bulges/internal loops or have a maximal number of helices in the exterior loop.

Next, we count the number of lattice points in the interior of each face of $P_n$.  For an edge of the form $E=\mathrm{conv}\{(x_1,y_1,z_1),(x_2,y_2,z_2)\}$, we use the formula
$$\#\left(\mathrm{int}E \cap \mathbb{Z}^3\right) =\gcd{\left(|x_1-x_2|,|y_1-y_2|,|z_1-z_2|\right)} -1$$ and obtain the following counts.  The edges $\mathrm{conv}\{(n,n,0), (1,\frac{n+1}{2},0)\}$ and $\mathrm{conv}\{(1,1,n-1), (1,\frac{n+1}{2},0)\}$ each have $\frac{1}{2}(n-3)$ lattice points in their interiors.  A total of $\frac{1}{2}(n-5)$ lattice points are in the interior of $\mathrm{conv}\{(1,n-1,0), (1,\frac{n+1}{2},0)\}$.  The interior of $\mathrm{conv}\{(n,n,0),(1,1,n-1)\}$ contains $n-2$ lattice points, and there are no interior lattice points for the edges $\mathrm{conv}\{(n,n,0), (1,n-1,0)\}$ and $\mathrm{conv}\{(1,1,n-1), (1,n-1,0)\}$.   

To determine the number of lattice points in a facet $F$ of $P_n$, we use Pick's theorem \cite{pickthm} $$\# \left(\mathrm{int}F\cap \mathbb{Z}^3\right)=\mathrm{Area}(F)-\frac{1}{2}\left[\#\left(\partial F\cap \mathbb{Z}^3\right)\right] +1,$$ where the area of $F$ is normalized with respect to the $2$-dimensional sublattice containing $F$.  We illustrate Pick's theorem with the following proposition.
\begin{proposition}\label{pick}
There are no interior lattice points in the facet $$F=\mathrm{conv}\{(1,1,n-1),(1,n-1,0), (n,n,0)\}.$$
\end{proposition}
\proof
 The triangle $F$ lies on the hyperplane $-X+(n-1)Y+(n-2)Z=n^2-2n$ in $\mathbb{R}^3$, and thus we normalize the area of $F$ by dividing by $\sqrt{(-1)^2+(n-1)^2+(n-2)^2}=\sqrt{2(n^2-3n+3)}$.  Before normalization, the area of $F$ is
\begin{eqnarray*}
&& \frac{1}{2}\sqrt{\left|\begin{matrix}1&1&n\\1&n-1&n\\1&1&1 \end{matrix}\right|^2 +\left|\begin{matrix}1&n-1&n\\n-1&0&0 \\1&1&1 \end{matrix}\right|^2 +\left|\begin{matrix}n-1&0&0\\1&1&n\\1&1&1 \end{matrix}\right|^2} \\ && =\frac{1}{2}\sqrt{2n^4-10n^3+20n-18n+6} \\ && =\frac{1}{2}(n-1)\sqrt{2(n^2-3n+3)}.
\end{eqnarray*}
Moreover, using the counts above for the interior lattice points in the edges of $F$, we have $$\#\left(\partial F\cap \mathbb{Z}^3\right)= (n-2)+0+0+3=n+1.$$
Applying Pick's theorem yields
\begin{eqnarray*}
 \# \left(\mathrm{int}F\cap \mathbb{Z}^3\right)&&=\textstyle{\frac{1}{2}}(n-1)-\textstyle{\frac{1}{2}}(n+1)+1 \\&&=0.
\end{eqnarray*}
\qed

For the other three facets of $P_n$, each contains $\frac{1}{4}(n-3)^2$ interior lattice points.  In total, this gives $\frac{1}{4}(3n^2-8n+13)$ lattice points on $\partial P_n$, all of which correspond to plane trees.

\subsection{Biological meaning of $P_n$ and $\mathcal{N}(P_n)$}
\subsubsection{The vertices of $P_n$}  The vertices of $P_n$ represent the secondary structures with the maximum number of helices in a loop---so-called ``maximal degree of branching''---and the fewest helices in a loop---or ``minimal degree of branching''---as described below.

If $T$ is a plane tree represented as a vertex of $P_n$ then $T$ has $n$ edges and $n+1$ vertices.  If in addition, $T$ has count vector $(n,n,0)$ then the degree of the root vertex is $n$ and the $n+1$ vertices are the root together with the $n$ leaves (vertices with $0$ children).  Thus, a secondary structure corresponding to $T$ has no internal loops, bulges, or multi-branch loops and the exterior loop has $n$ helices. 

If $T$ has count vector $(1,1,n-1)$, the root vertex has degree $1$, there is one leaf, and $n-1$ vertices of degree $2$ ($1$ child).  Thus, $T$ is a straight line, and a secondary  structure corresponding to $T$ has no multi-branch loops and the exterior loop has one helix..  

If $T$ has count vector $(1,n-1,0)$, the $n+1$ vertices are the root (with degree $1$), $n-1$ leaves, and one vertex of degree $n$.  Secondary structures corresponding to $T$ have no internal loops or bulges and one multi-branch loop with $n$ helices.  In addition, the exterior loop has one helix.

The remaining vertices---$(1,\frac{n+2}{2},0)$ for $n$ odd and $(1,\frac{n}{2},1)$, $(2,\frac{n+2}{2},0)$, or  $(1,\frac{n+2}{2},0)$ for $n$ even---are dealt with in the following proposition.
\begin{proposition} \label{vertbranch}
\begin{enumerate}[(i)] 
\item  For $n$ odd, any plane tree with count vector $(1,\frac{n+1}{2},0)$ satisfies $d_2=\frac{n-1}{2}$ and $d_i=0$ for $i > 2$.  \label{nodd}
\item For $n$ even, any plane tree with count vector $(1,\frac{n}{2},1)$ or $(2,\frac{n+2}{2},0)$ satisfies $d_2=\frac{n-2}{2}$ and $d_i=0$ for $i> 2$. \label{neven1}
\item For $n$ even, any plane tree with count vector $(1,\frac{n+2}{2},0)$ satisfies $d_2 = \frac{n-4}{2}, d_3=1,$ and $d_i=0$ for $i>3$. \label{neven2}
\end{enumerate} 
\end{proposition}
\proof
For \eqref{nodd}, suppose $n$ is odd and $T$ is a plane tree with $n$ edges, $r=1$, $d_0=\frac{n+1}{2}$, and $d_1=0$.  Then,  $T$ has $\frac{n+1}{2}+1$ vertices of degree $1$, and the remaining $n+1-\left(\frac{n+1}{2}+1\right)=\frac{n-1}{2}$ vertices have degree at least $3$.  Thus,
\begin{eqnarray*}
\sum_{v\in V}\mathrm{deg}v && =\textstyle{\frac{1}{2}}(n+1)+1+\displaystyle{\sum_{\mathrm{deg}v\, \ge \, 3}\mathrm{deg}v} \\
&&\ge  \textstyle{\frac{1}{2}}(n+1)+1+ \textstyle{\frac{3}{2}}(n-1) \\
&&=2n
\end{eqnarray*}  However, since $\displaystyle{\sum_{v\in V}\mathrm{deg}v=2|E|}$, we must have equality.  Thus, all other vertices must have degree $3$ ($2$ children).

The proof of \eqref{neven1} is nearly identical to that of \eqref{nodd}.

For \eqref{neven2}, a plane tree with $n$ edges, $r=1$, $d_0=\frac{n+2}{2}$ and $d_1=0$ has $\frac{n+4}{2}$ vertices of degree $1$ and zero vertices of degree $2$.  Such a tree cannot have all other vertices of degree $3$ as this would yield a graph with an odd number of odd vertices.  Thus, there is a vertex $v_0$ with degree $p$ with $p\ge 4$ even.  This gives
\begin{eqnarray*}
\sum_{v\in V}\mathrm{deg}v && =\textstyle{\frac{1}{2}}(n+4)+p+\displaystyle{\sum_{{\scriptsize \begin{array}{c}\mathrm{deg}v\ge 3\\ v\neq v_0 \end{array}}} \mathrm{deg}v} \\
&&\ge  \textstyle{\frac{1}{2}}(n+4)+4+ \textstyle{\frac{3}{2}}(n-4) \\
&&=2n
\end{eqnarray*}
As before this inequality must be an equality, and hence $p = 4$ and all other vertices have degree $3$.
\qed

\begin{figure}[htbp]
\begin{center}
\includegraphics[width=5in]{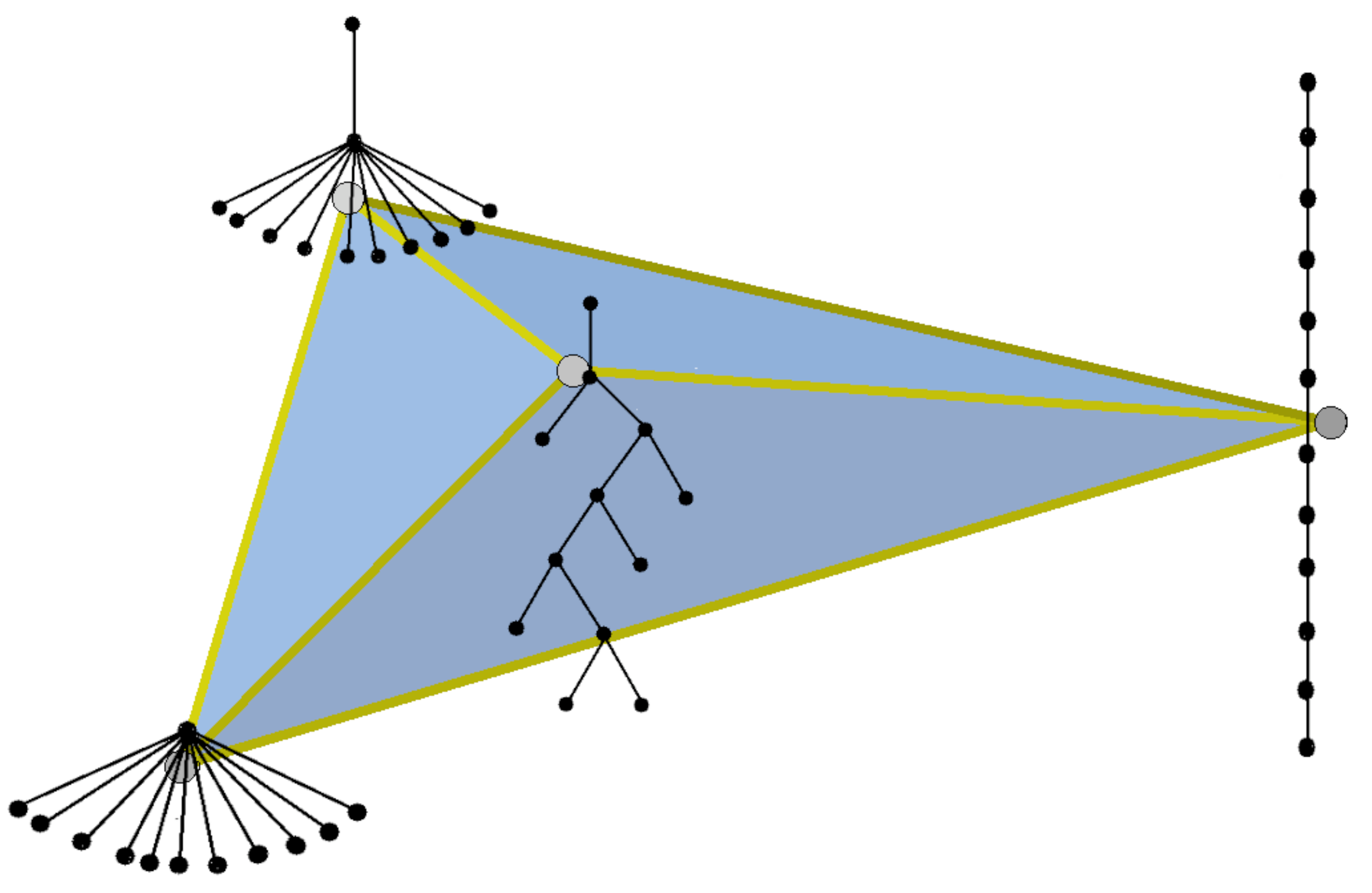}
\caption{The RNA polytope $P_n$.}
\label{figure:tope}
\end{center}
\end{figure}

Thus, for $n$ odd, the count vector $(1,\frac{n+1}{2},0)$ corresponds to secondary structures with no interior loops/bulges, all multi-branch loops have $3$ helices, and the exterior loop has one helix.  When $n$ is even, a secondary structure with $n$ helices and all three of these properties is not possible.  We instead have three cases, each with exactly one of the properties relaxed: a structure corresponding to $(1,\frac{n}{2},1)$ has one interior loop/bulge, the count vector $(1,\frac{n+2}{2},0)$ arises from structures having one multi-branch loop with $4$ helices (all other multi-branch loops have $3$ helices), and the exterior loop of a structure corresponding to $(2, \frac{n+2}{2},0)$ has $2$ helices.  For $n$ odd, plane trees representative of those described in this section are shown in Figure \ref{figure:tope}.
\begin{remark} The map from plane trees to count vectors is generically many-to-one.  Three of the $4$ vertices, however, correspond to exactly one tree: $(n,n,0), (1,1,n-1), (1,n-1,0)$.   The trees with count vector $(1, \frac{n+1}{2},0)$ are in one-to-one correspondence with full binary trees with $n-1$ edges (by removing the root vertex).  There are $C_{\frac{n-1}{2}}$ such trees \cite{aerated}, where $C_{\frac{n-1}{2}}$ is the $\frac{n-1}{2}$th Catalan number defined in equation \eqref{catalan}.
\end{remark}
\subsubsection{The rays in $\mathcal{N}(P_n)$} 
The energy function $E^{\prime}$  in equation \eqref{eprime} scores a secondary structure with $n$ helices based on the number of helices in the exterior loop, the number hairpin loops, and the number of bulges/internal loops.  The normal fan $\mathcal{N}(P_n)$ of $P_n$ subdivides the $(\theta_2, \theta_3, \theta_4)$ parameter space.  Each vector in $(x,y,z)\in \mathbb{R}[\theta_2]\times\mathbb{R}[\theta_3]\times \mathbb{R}[\theta_4]$ corresponds to a scoring function in which $x$ gives the weight of a helix in the external loop, $y$ gives the weight of a hairpin loop, and $z$ gives the weight of a bulge/internal loop.  

The fan $\mathcal{N}(P_n)$ consists of cones generated by elements in the power set  
$$\mathscr{P}\left(\{(1,0,0),(0,0,1),(-1,2,1),(1,1-n,2-n)\} \right).$$  Thus, a parameter vector $v \in  \mathbb{R}[\theta_2]\times\mathbb{R}[\theta_3]\times \mathbb{R}[\theta_4]$ has the form $c_1 y_1 +c_2y_2+ c_3y_3$ with $c_1,c_2,c_3\ge 0$ and $y_1,y_2,y_3 \in \{(1,0,0),(0,0,1),(-1,2,1),(1,1-n,2-n)\}$.  A generic vector in $\mathbb{R}^3$ lies in the interior of one of the $3$-dimensional cones in $\mathcal{N}(P_n)$, and hence we give a brief interpretation of the parameter vectors with $c_i\neq 0$ for $i=1,2,3$.

Scoring vectors in the interior of the cone $\langle (0,0,1),(1,0,0),(1,1-n,2-n)\rangle$ penalize for hairpin loops and can independently penalize or reward for helices in the exterior loop and bulges/internal loops.  If $v \in \mathrm{int}\langle (0,0,1),(1,0,0),(-1,2,1)\rangle$ then $v$ gives a penalty for both hairpin loops and interior loops/bulges.  Helices in the exterior loop can be beneficial or harmful with this scoring vector, and $v$ can equally penalize helices in the exterior loop, hairpin loops, and internal loops/bulges.   Scoring vectors in the interior of one of the two remaining cones can reward or penalize all three quantities.  These are not independent,  however.  For instance, if $v \in \mathrm{int}\langle (1,1-n,2-n),(0,0,1),(-1,2,1)\rangle$ and hairpin loops are disadvantageous under $v$'s scoring scheme then helices in the exterior loop are beneficial.  If  $w\in  \mathrm{int}\langle (1,0,0),(1,1-n,2-n),(-1,2,1)\rangle$ and $w$ rewards hairpin loops then $w$ rewards bulges/internal loops.  Similarly, if $w$ penalizes for bulges/internal loops then $w$ penalizes for hairpin loops.  Also, scoring vectors in the interior of the cone $\langle (1,1-n,2-n),(0,0,1),(-1,2,1)\rangle$  can equally reward hairpin loops, internal loops/bulges, and helices in the exterior loop. 

\subsection{Variation in the parameter space}
In this section, we add additional information to the parameters $\{\theta_2,\theta_3,\theta_4\}$ in order to study the effect of varying the multi-branch loop parameters in the thermodynamic model of RNA folding. 
We obtain free energy parameters for plane trees using one of the  four combinatorial sequences having the form $$X^4(Y^6X^4Z^6X^4)^k\quad \mbox{where} \quad k\ge1 \quad \mbox{and}\quad \left\{ \begin{matrix}  X=A & \mbox{and} &\{Y,Z\}=\{C,G\}\\  X=C & \mbox{and} &\{Y,Z\}=\{A,U\} \end{matrix}\right. . $$  
In these sequences, the segments of the form $Y^6$ pair with the $Z^6$ segments while the $X$ nucleotides remain unpaired, and moreover all the loops of a given type have the same free energy.  We do not include the possibilities $X=U$ and $\{Y,Z\}=\{C,G\}$ or $X=G$ and $\{Y,Z\}=\{A,U\}$ because we want to prevent the $G-U$ pairing.   For a given sequence, we use both the current (version 3.0) \cite{mathews-etal-04} and previous (version 2.3) \cite{walter-turner-94} thermodynamic parameters, determined by the Turner lab.  The parameters $a_3, a_0,$ and $a_1$ are based on experimental measurement and are listed in Table \ref{table:param}.  The parameters $a_2$ and $c_2$ come from the multi-branch loop scoring function in equation \eqref{30}, where the parameters $a,b$, and $c$ in this function are not determined experimentally.  If $L$ is a multi-branch loop with $n_1$ single-stranded bases and $n_2$  helices and $L$ appears in a secondary structure for one of our $4$ combinatorial sequences, then we have $n_1=4n_2$. Additionally, for each helix in $L$, the single-base stacking energy is $a_3$.  Thus, free energy of $L$ in equation \eqref{30} becomes $E(L)=a+ 4bn_2+cn_2+a_3n_2$, and the parameters $a_2$ and $c_2$ in the free energy function $E^{\prime}$ in equation \eqref{eprime} can be written as $a_2=4b+c+a_3$ and $c_2=a$.
\begin{table}[htdp] 
\centering      
\begin{tabular}{c c c c | c c c }  
\hline\hline                         
&\multicolumn{3}{c}{Turner 3.0 Values}  &  \multicolumn{3}{c}{Turner 2.3 Values} \\  [-1ex]  \raisebox{1.5ex}{Sequence} & $ a_3 $& $ a_0 $& $ a_1$  &  $ a_3$ & $ a_0$ & $a_1$ 
\\ [0.5ex]    
\hline
X=A, Y=G, Z=C & $-1.9$ & $4.1$ & $2.3$   & $-1.9 $ & $3.5$ & $3.0$ \\
X=A, Y=C, Z=G & $-1.6$ & $4.5$ & $2.3$  & $-1.6 $ & $3.8 $ & $3.0$  \\ 
X=C, Y=A, Z=U  & $-0.4$ & $5.0$ & $3.7$ & $-0.4 $ & $4.3 $ & $4.0$  \\
X=C, Y=U, Z=A  & $-0.6$ & $4.9$ & $3.7$ & $-0.6$ & $4.2 $ & $4.0$ \\
\hline
\end{tabular} 
\caption{Energy parameters for plane trees}
\label{table:param} 
\end{table}
Table \ref{table:variation} illustrates three types of variation: variation of specific nucleotides in combinatorial sequence, variation of the version of Turner's  energy parameters, and variation of $a,b,c$ parameters.    The effect of varying the multi-branch loop parameters $a,b,c$ is more or less the same for each sequence and energy table: two different count vectors can be minimal depending on the value of $a+12b+3c$.  Technically, a third vertex of $P_n$ has minimal energy in some cases when $b=c=0$.  However, if the offset and helix penalties are both zero, the multi-branch energy function will have no penalties for the number of single-stranded bases and the number of stems in a loop.  This does not agree with the free energy model.  

Varying the sequence alone, we obtain differences in the cut-off values for $a+12b+3c$.  On the whole, however, nucleotide variation in the combinatorial sequence does not give qualitative differences in the minimal energy plane trees.  

We do see (in $3$ of the $4$ sequences) qualitative differences in the minimal energy trees when we compare version 3.0 parameters to version 2.3 parameters.  For instance, when $a+12b+3c$ is large, $3$ of the $4$ sequences give the `straight line' tree with count vector $(1,1,n-1)$ minimal with version 3.0 parameters.  Using version 2.3, all four sequences result in the maximal degree of branching, with count vector $(n,n,0)$ having minimal energy.  This difference in minimal energy trees is not too surprising because the change from version 2.3 to version 3.0 was based on more accurate experimental measurement.  The secondary predicted structures have indeed changed.   
\begin{table}[ht] 
\centering      
\begin{tabular}{c c c c c}  
\hline\hline                         
Vertex& Rays in $\mathcal{N}(P_n)$ &Energy version & Restrictions on $a,b,c$& $\begin{matrix}\mbox{Sequence: } \\ [\mathrm{X,Y,Z}] = \end{matrix}$ \\  \hline
$(1,1,n-1)$&$\begin{matrix} (1,1-n,2-n) \\ (1,0,0)\\ (-1,2,1) \end{matrix}$ & $3.0\quad (2.3) $ & $ a+12b+3c \ge \left\{\begin{matrix}\mbox{N/A}&(\mbox{N/A}) \\4.9&(\mbox{N/A}) \\3.6 &(\mbox{N/A})\\ 4.3 &(\mbox{N/A})\end{matrix} \right. $ &$\begin{matrix} [\mathrm{A,G,C}]  \\ [\mathrm{A, C, G}]  \\ [\mathrm{C, A, U}]\\  [\mathrm{C, U, A}]  \end{matrix} $ \\   \\
$(1,\frac{n+1}{2},0)$ & $\begin{matrix} (0,0,1) \\ (1,0,0)\\ (-1,2,1) \end{matrix}$& $3.0\quad (2.3) $ & $ a+12b+3c \le \left\{\begin{matrix}6.0 & (5.4) \\4.9&(5.4) \\3.6&(4.7) \\ 4.3&(4.8) \end{matrix} \right. $ &$\begin{matrix} [\mathrm{A,G,C}]  \\ [\mathrm{A, C, G}]  \\ [\mathrm{C, A, U}]\\  [\mathrm{C, U, A}]  \end{matrix} $  \\   \\
$(n,n,0)$& $\begin{matrix} (1,1-n,2-n) \\ (0,0,0)\\ (-1,2,1) \end{matrix}$& $3.0\quad (2.3) $ & $ a+12b+3c \ge \left\{\begin{matrix}6.0 & (5.4) \\ \mbox{N/A} &(5.4) \\  \mbox{N/A}&(4.7) \\ \mbox{N/A}&(4.8) \end{matrix} \right. $ &$\begin{matrix} [\mathrm{A,G,C}]  \\ [\mathrm{A, C, G}]  \\ [\mathrm{C, A, U}]\\  [\mathrm{C, U, A}]  \end{matrix} $  \\   \\
$(1,n-1,0)$&$\begin{matrix} (0,0,1) \\ (1,0,0)\\ (1,1-n,2-n) \end{matrix}$ & $3.0\quad (2.3) $ & $b=c=0, a = \left\{\begin{matrix}6.0 & (5.4) \\ \mbox{N/A} &(5.4) \\  \mbox{N/A}&(4.7) \\ \mbox{N/A}&(4.8) \end{matrix} \right. $ &$\begin{matrix} [\mathrm{A,G,C}]  \\ [\mathrm{A, C, G}]  \\ [\mathrm{C, A, U}]\\  [\mathrm{C, U, A}]  \end{matrix} $  \\  
\hline
\end{tabular} 
\caption{Restrictions on $a,b,c$ parameters from full-dimensional cones in $\mathcal{N}(P_n)$}
\label{table:variation} 
\end{table} 
It is worth noting that if we use the actual penalties for offset, free base, and helix from versions $2.3$ and $3.0$ of the Turner energies, we obtain
$$a+12b+3c = \left\{\begin{array}{cc}4.6 & v3.0 \\ 9.7 & v2.3 .\end{array}\right.$$
Thus, all four combinatorial sequences yield $(n,n,0)$ with minimal energy for version $2.3$.  Moreover, as 9.7 is a fair amount greater than the cut off for all four sequences, slight variation in these parameters will not change the predicted structure.  For version $3.0$, the two combinatorial sequences with unpaired poly-A segments have $(1,\frac{n+1}{2},0)$ being minimal while $(1,1,n-1)$ is minimal for the other two sequences.  Also, 4.6 is much closer to the cut off values for the sequences.  Small changes in these parameters could change which trees have minimal energy. 
\subsection{RNA STRAND database analysis}
\subsubsection{Overall shape of data}
Our initial collection of secondary structures contains $145$ structures from $137$ distinct RNA sequences, as described in Materials and Methods.  The sequences range from $19$ to $4216$ nucleotides.  We exclude structures for which the number of helices is less than $5$ from further analysis.  This reason for this is that not all the vertices of $P_n$ listed in Proposition \ref{psin} are valid and distinct when $n\leq 4$. We have $110$ structures with $n\ge 5$ (from $103$ sequences) having average (median) length of $ 739\; (367)$ and $n = 27\: (13)$.  We break these into classes, based on the number of helices, as depicted in Table \ref{table:category}.    

While our collection contains more small and medium trees as compared to large trees, this reflects the frequency in the RNA STRAND database.  For instance, according to an analysis done by RNA STRAND,  the average (median) number of helices over the entire database is $28 \: (8)$.  This count does, however, include  the sequences with fewer than $5$ helices and includes a less restrictive definition of bulges/internal loops and helices: internal loops/bulges can have any number of unpaired bases and helices can have any number of base pairs.  Our large trees come from $16$S ribosomal RNA and $23$S ribosomal RNA sequences and have a minimum sequence length of  $954$ nucleotides.  In the RNA STRAND database, only $20$\% of the $4666$ structures contain at least $954$ nucleotides.   

\begin{table}[htdp] 
\centering      
\begin{tabular}{c c c c c c c}  
\hline\hline                         
 Category  &Range of $n$ &\# of trees &Average length&Median length& Average $n$ &  Median $n$ 
\\ [0.5ex]    
\hline
Small  & $5-12$ & $50$&$244$ & $220$ & $9$ & $9$  \\
Medium & $13-40$ & $40$ & $676$ & $512$ & $22$ & $19$ \\
Large & $41-136$ & $20$ & $2104$ & $1831$ & $82$ & $76$  \\
\hline
\end{tabular} 
\caption{Trees in RNA STRAND collection by size}
\label{table:category} 
\end{table}

\subsubsection{Location of count vectors on polytope}
It is of great importance to know when biologically correct secondary structures can be predicted by the free energy model.  With our simplified energy function $E^{\prime}$ in \eqref{eprime}, we ask if the biologically correct structures can be minimal for some choice of parameters.  As mentioned in \textsection \ref{boundsec}, this translates into determining when the corresponding count vectors lie on the boundary of $P_n$.  

\begin{figure}[htdp]
\begin{center}
\includegraphics[width=4.5in]{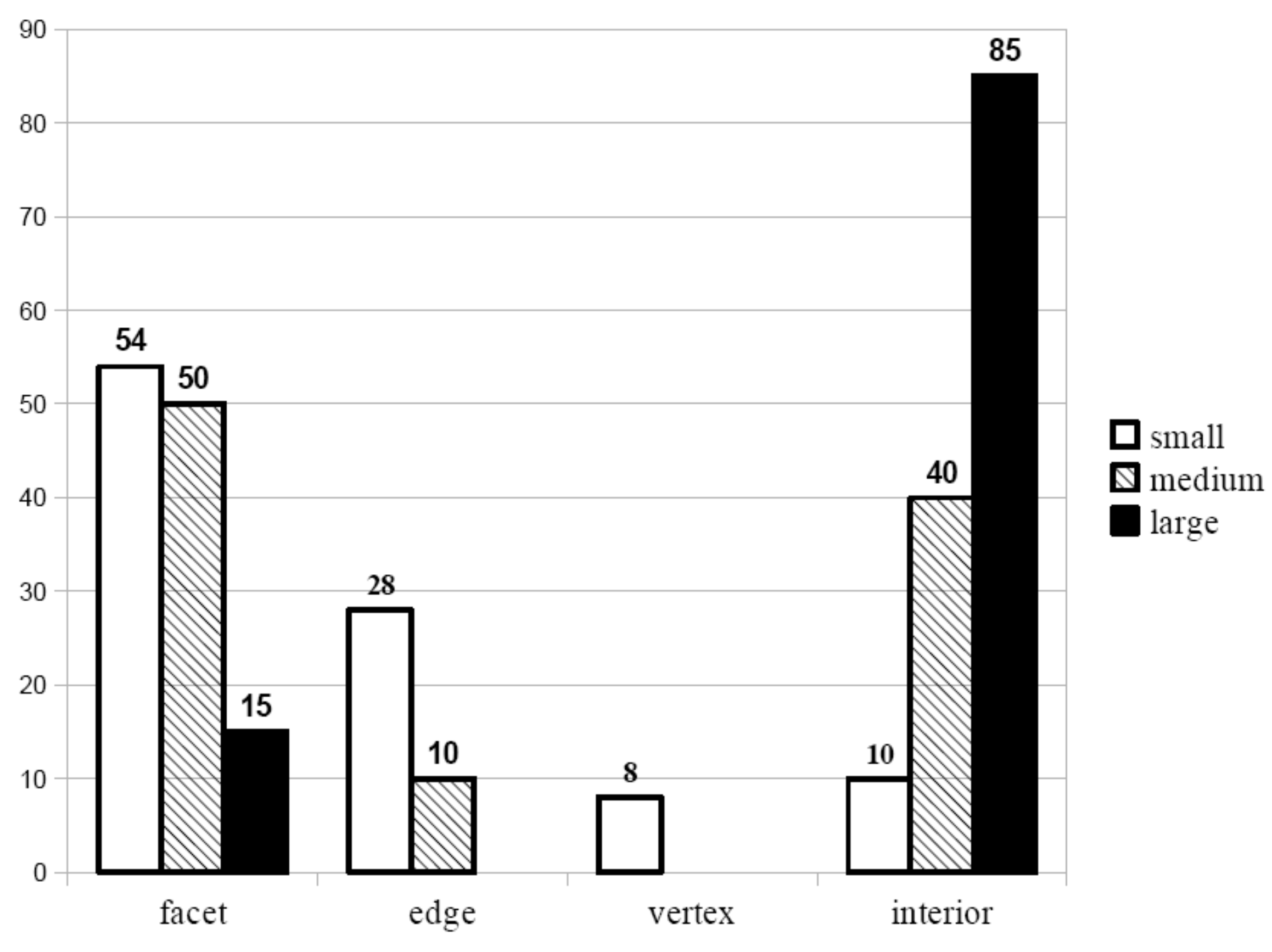}
\end{center}
\caption{Location of count vectors on $P_n$ for small, medium, and large trees (in percentage)}
\label{figure:location}
\end{figure}

Seventy-one out of 110 count vectors lie on the boundary of $P_n$: $49$ lying on the interior of a facet, $18$ lying on the interior of an edge, and $4$ occurring as vertices.  The average number of edges for plane trees on the boundary of $P_n$ is  $17$ and is $45$ for plane trees in the interior of $P_n$.  Of those contained in the interior of a facet, $28$ are minimal for parameters in $\langle(-1,2,1)\rangle$, $7$ are minimal for parameters in $\langle(0,0,1)\rangle$, and $14$ are minimal for parameter values in $\langle(1,0,0)\rangle$.  Of those contained in the interior of an edge, $8$ are minimal for parameters in $\langle(-1,2,1), (1,0,0)\rangle$, $9$ are minimal for parameters in $\langle(-1,2,1, (1,1-n,2-n)\rangle$, and $1$ is minimal for parameters in $\langle(1,0,0),(0,0,1)\rangle$.  The $4$ count vectors that are vertices of $P_n$ satisfy $n=5$ or $6$ and consist of the set $\{(5,5,0), (6,6,0), (1,4,0), (1,1,4)\}$.  Figure \ref{figure:location} shows the location of the count vectors for small, medium, and large trees, given in terms of the percentage of trees in each category.

\subsubsection{Closest vertex to count vectors}
In order to determine which of the $4$ vertices of $P_n$ is closest to a given count vector, we map the tetrahedron $$\mathrm{conv}\{(1,n,n,0), (1,1,1,n-1), (1,1,n-1,0), (1, 1, \textstyle{\frac{n+1}{2}}, 0)\}$$ onto the standard tetrahedron with vertices $\{(1,0,0,0), (0,1,0,0), (0,0,1,0), (0,0,0,1)\}$.  This is accomplished with the following matrix  
\begin{equation} \label{matrix}\left[\begin{matrix}\displaystyle{ -\frac{1}{n-1}}&\displaystyle{ \frac{1}{n-1}}& 0&0 \\ \\ 0&0&0&\displaystyle{ \frac{1}{n-1}} \\ \\ \displaystyle{ -\frac{n}{n-3}}&\displaystyle{ -\frac{1}{n-3}}&\displaystyle{ \frac{2}{n-3}}&\displaystyle{ \frac{1}{n-3} }\\ \\ \displaystyle{ \frac{2n(n-2)}{(n-1)(n-3)} }&\displaystyle{ \frac{2}{(n-1)(n-3)}}& \displaystyle{ -\frac{2}{n-3}} &\displaystyle{ -\frac{2 (n-2)}{(n-1)(n-3)}}
\end{matrix}\right] \end{equation}
which has determinant $\frac{2}{(n-1)^2(n-3)}$.  For a given $n$, any count vector $(r,d_0,d_1)$ can be written as a sum
$$a_1(n,n,0) + a_2(1,1,n-1) + a_3(1,n-1,0) + a_4(1, \textstyle{\frac{n+1}{2}}, 0)$$ with $0\leq a_i \leq 1$ and $a_1 + a_2 + a_3 + a_4 =1$.  After applying the linear transformation \eqref{matrix}, the lattice point $(1,r,d_0,d_1)$ will have coordinates $(a_1,a_2,a_3,a_4)$.  The coordinate $a_i$ gives a measure of the `closeness' to vertex $i$.  For a given RNA structure, the largest of the $a_i$ gives the vertex closest to its count vector.  Moreover, if $t=\mathrm{max}\{a_1, a_2, a_3, a_4\}$ then $0.25 \leq t \leq 1$.

Fifty-two of the $110$ structures are closest to $(1, \frac{n+1}{2}, 0)$, $38$ are closest to $(1, 1, n-1)$, $10$ are closest to $(n, n, 0)$, and $6$ are closest to $(1, n-1, 0)$.  Additionally, we have $2$ that are closest to both $(1, 1, n-1)$ and $(n, n, 0)$ and $2$ that are closest to both $(1, 1, n-1)$ and $(1, \frac{n+1}{2}, 0)$.  The average values of $(a_1, a_2, a_3, a_4)$ over the $110$ structures are $(0.181, 0.357, 0.138, 0.332)$  which shows that as a whole, the count vectors are closest to $(1, 1, n-1)$ and $(1, \frac{n+1}{2}, 0)$.

We say a count vector is `close' to vertex $i$ if $a_i > 0.625.$  The value $0.625$ is halfway in between the smallest and largest possible values of $a_i$.  With this definition, $22\%$ of the small trees are close to vertices, $5\%$ of the medium trees are close to vertices, and no large trees are close to vertices.  Thirteen trees in total are close to vertices, of which $8$ are close to $(1,1,n-1)$, $ 2$ are close to $(1, \frac{n+1}{2},0)$, $2 $ are close to $(n,n,0)$, and $ 1$ is close to $(1,n-1,0)$.  All thirteen of these lattice points lie on the boundary of $P_n$ and hence correspond to minimal energy trees for some choice of parameter values.    
\section{Discussion and Conclusions}
We have used a simple scoring scheme for scoring RNA folds: energy is assigned to a secondary structure based solely on the total number of helices, the number of helices in the exterior loop, and the numbers of hairpin loops and bulges/internal loops.  Fixing the total number of helices, the extremal folds are those with the maximal and minimal degrees of branching.   When a generic parameter vector is chosen, precisely one of those will have minimal energy.  For more specific choices of parameters (biologically realistic or not), the number of minimal count vectors is on the order of the square of the total number of helices.  While this seems large, the total number of count vectors that cannot be minimal for any choice of parameters is on the order of the cube of the total number of helices.  Thus, when this total is large, we would not expect such a scoring scheme to accurately predict the correct structures.  This is supported by our RNA STRAND analysis in which $85\%$ of the count vectors from known structures with a high number of helices cannot be minimal for any choice of parameters.  None of these structures are `close' to the extremal folds.  This is not unexpected, however, since even the highly detailed free energy model is not accurate for large RNA molecules \cite{doshi}.

On the other hand, when the total number of helices is small, only $10\%$ of the known structures cannot be minimal for our scoring scheme.  While the scoring function used in this work is too simplistic to implement in a prediction software, our results suggest that for small RNA molecules, the full free energy model is not necessary for accurate predictions.  We are not the first to make this observation, for \cite{stoch} analyzed some simple probabilistic RNA folding models---one with as few $21$ free parameters---whose accuracies are comparable to mfold's.  In their study, the sequences used for testing came from ribonuclease P RNA, transfer mRNA, and signal recognition particle RNA sequences, all of which yield small to medium trees by our classification.  While $21$ parameters is far too many for parametric analysis using polyhedral geometry, perhaps a simple model incorporating some thermodynamics and some probabilistic parameters can accurately predict the folding of small RNA molecules. 

We compared the variation of multi-branch loop parameters to two other types of variation in the parameter space.  Fixing the combinatorial sequence and energy version, two possible count vectors can be minimal by varying the multi-branch loops parameters.  If we use the most recent (accurate) energy version, we find that for 3 of the 4 sequences, these two count vectors include $(1,1,n-1)$ and $(1,\frac{n+1}{2},0)$.  Interestingly, these two vertices are closest to the known structures in our RNA STRAND collection.  Moreover, regardless of the choices of multi-branch loop parameters in the current version of the thermodynamic model, predicted structures have a low degree of branching---both in the exterior loop and in the multi-branch loops.  Out of the three possible variations, the most significant changes come from varying the energy version, as the possible predicted structure for version 2.3 have a high degree of branching.  Even though the penalties for off-set, free base and helix in the multi-branch loop energy calculation are chosen without specific measurement, they do not appear to have a dramatic effect on the predicted structures.  One would hope that the parameters determined experimentally are what truly govern the predicted structures, and our findings support this possibility. 
\section{Materials and Methods}
\subsection{Selection of secondary structures from RNA STRAND database}
The RNA STRAND database \cite{rnastrand} was searched by type of RNA (for example, $16$S ribosomal RNA, cis-regulatory element, or group I intron).  Each type of RNA was sorted by molecule length, and structures were selected from a variety of organisms to be representative of the different lengths appearing in the database for that type of RNA.  Visual inspection of the secondary structures was important in the selection of the structures for our collection.  It allowed for the inclusion of similar length structures with different types of branching.  It also prevented our collection from containing nearly identical structures formed by two different RNA molecules of the same type.  Finally, visual inspection kept our collection from having a plethora of structures with only one or two helices; these structures are overrepresented in the RNA STRAND database.
\subsection{Removal of pseudoknots from .ct files}
In order to obtain a plane tree from a give secondary structure, pseudoknots were removed.  A perl script read the .ct file and stored the closing pairs of all helices, where the helices are defined is \textsection \ref{calchelix}.  Each pair $(i,j)$ and $(i^{\prime}, j^{\prime})$ of closing pairs was tested to see if $i<i^{\prime}<j<j^{\prime}$.  If true, the pairs  $(i,j)$ and $(i^{\prime}, j^{\prime})$ were printed to a file.  Next, for each pair  $(i,j)$ and $(i^{\prime}, j^{\prime})$ in the output file, one of the associated helices was removed according to the following rubric.  If some closing pair $(i,j)$ appears multiple times, its helix was removed under the assumption that it formed a pseudoknot.  If both $(i,j)$ and $(i^{\prime}, j^{\prime})$ were not listed with any other closing pairs, the shorter of the $2$ corresponding helices was removed.  In the event that the two helices had the same number of paired bases, two versions of the .ct file were saved---one with the first helix removed and one with the second helix removed.
\subsection{Calculation of $n,r,d_0,d_1$ from .ct files} \label{calchelix}
After all the pseudoknots were removed from the .ct files of secondary structures in our collection, a perl script calculated $n,r,d_0,$ and $d_1$.  In our simplified model of RNA folding, all helices have the same energy independent of the number of base pairs in the helix.  Similarly, all bulges/internal loops have the same energy regardless of the number of free bases in the loop.  Because of this, very small bulges/internal loops and very short helices were ignored.  Bulges and interior loops were required to have at least $3$ unpaired bases.  No restrictions were placed on the number of free bases in a hairpin loop, which was important so as to maintain the graph structure (edges connecting two vertices).  
\begin{figure}[htbp]
\begin{center}
{\bf A)}\includegraphics[width=2.3in]{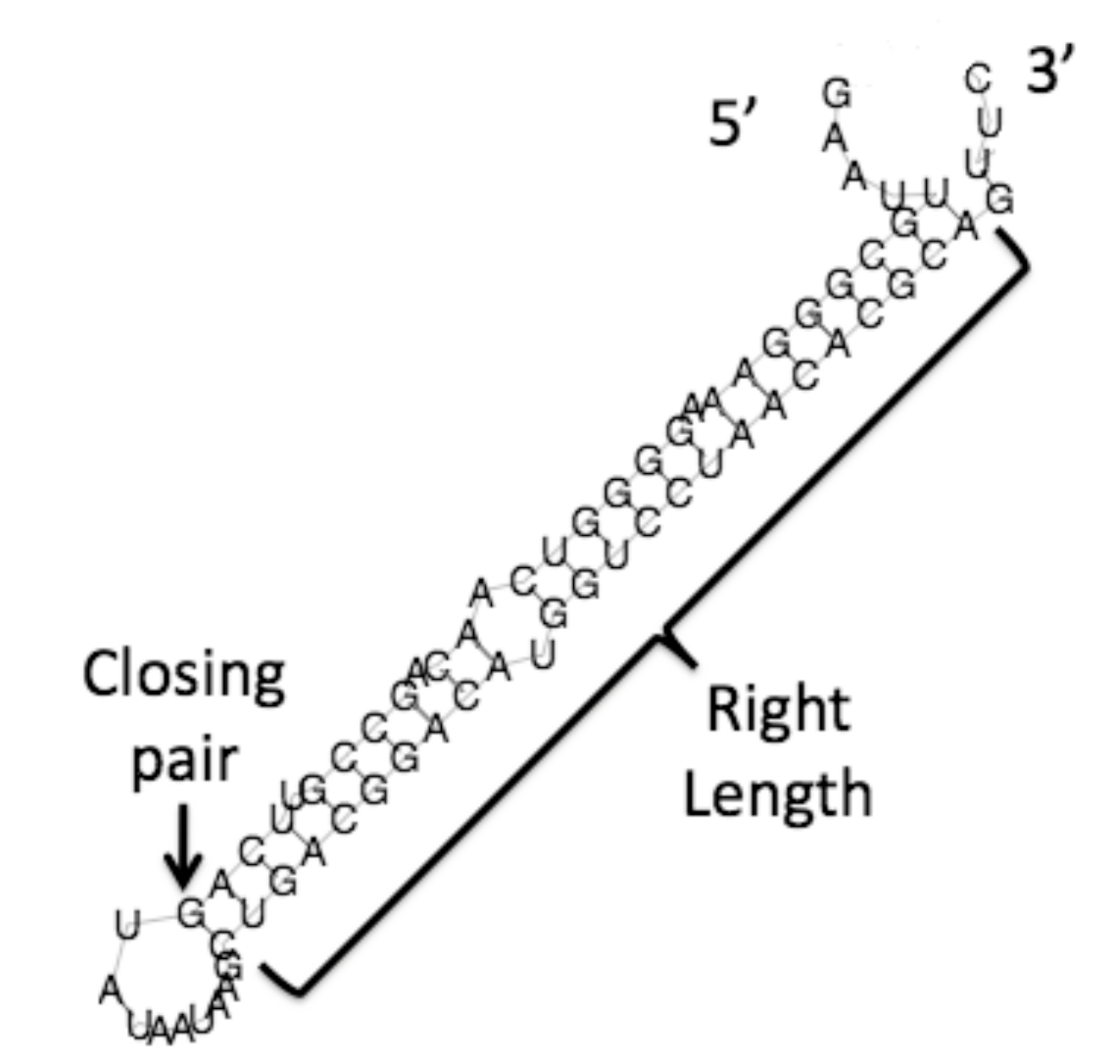}{\bf B)}\includegraphics[width=1.7in]{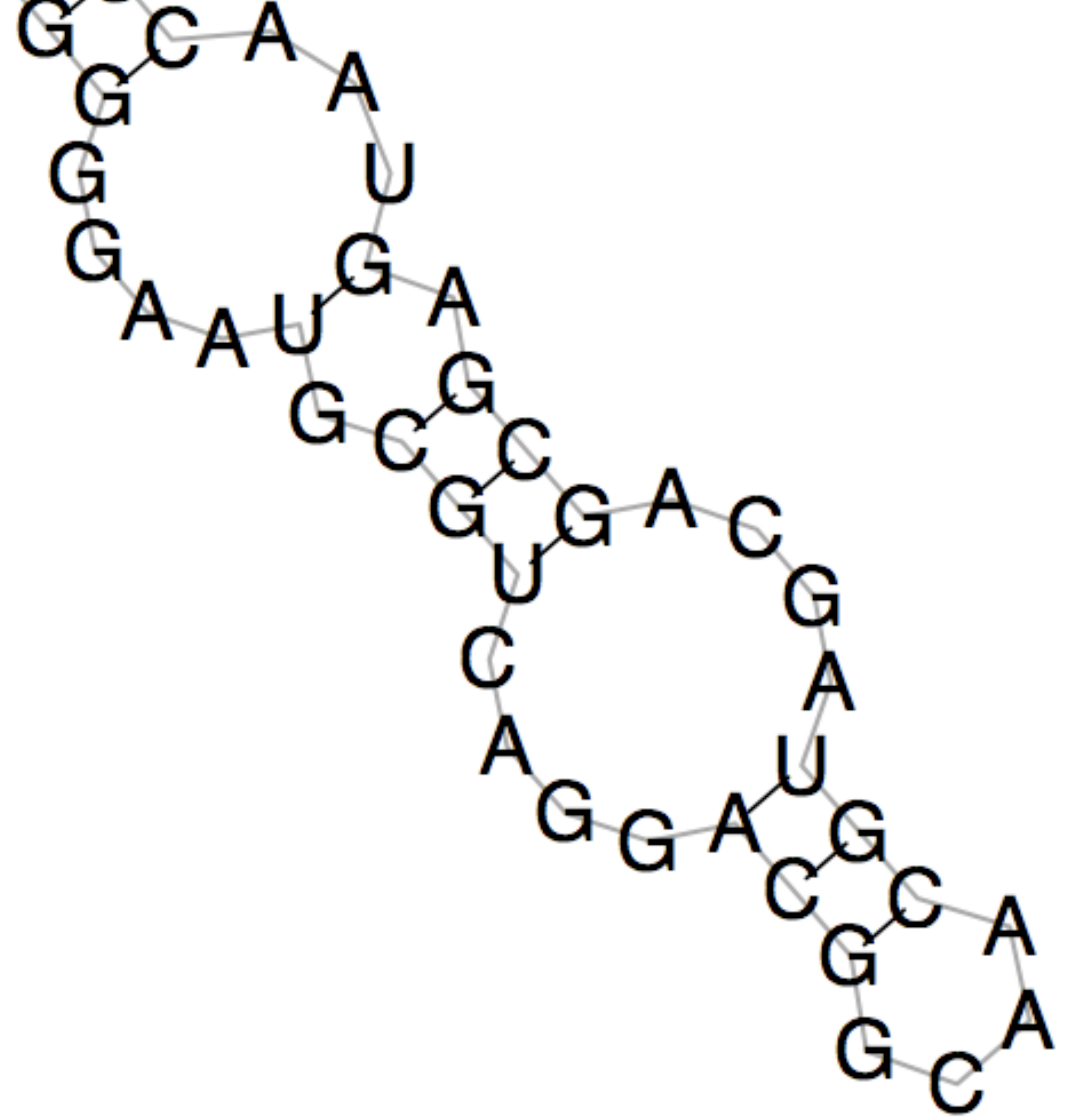}$\quad${\bf C)}$\quad$ \includegraphics[width=1.3in]{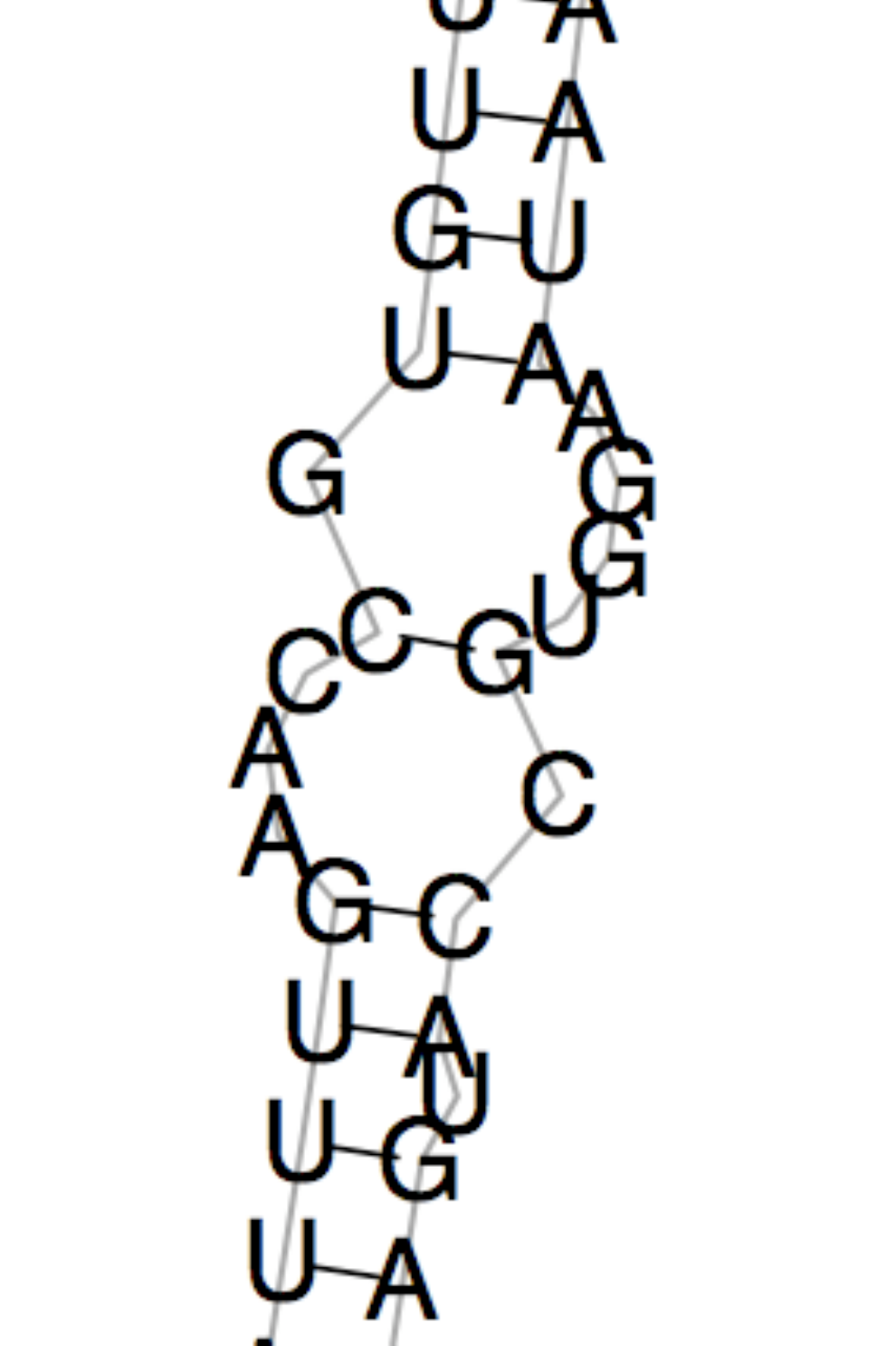}
\caption{Helices and internal loops: A, B, and C are fragments from structures in the RNA STRAND database}
\label{figure:helices}
\end{center}
\end{figure}

Each helix with choice of closing pair has a `left length' and `right length' of the helix.  The \emph{left length} of a helix is the number of bases in the portion of the sequence that terminates at one of the closing bases.  The \emph{right length} of a helix is the number of bases in the portion of the sequence that originates at one of the closing bases.  The closing pair of a helix as well as its right length are depicted in Figure \ref{figure:helices}{A}.  For this structure, the helix with closing pair G--C has left length $28$ and right length $25$.  For our analyses, a helix was defined to have both the left and right length $3$ or greater.  Thus, the piece of secondary structure shown in Figure \ref{figure:helices}{B} has two helices---one with left and right length $5$ and one with left and right length $3$---and one hairpin loop.  Similarly, with our definitions, the fragment depicted in Figure \ref{figure:helices}{C} has only $1$ interior loop that contains the base pairs G--C and U--A.   The single C--G base pair is not considered a helix, and since each of the internal loops containing the C--G pair have more than $3$ unpaired bases, the C--G base pair is not considered a part of either helix. 
\section*{Acknowledgements}
V.H. and C.E.H were both supported by the NIH grant 1R01GM083621-01 (P.I. Heitsch).  C.E.H. also acknowledges funding from a Career Award at the Scientific Interface (CASI) from the Burroughs
Wellcome Fund (BWF).  Additionally, V.H. would like to thank Justin Filoseta for the remarkable computer support at the Georgia Institute of Technology.
\bibliographystyle{abbrv}

 \end{document}